\pdfoutput=1

\documentclass[11pt]{article}

\usepackage{xspace}
\newcommand{\eat}[1]{}
\newcommand{\class}[1]{{\ensuremath{\mathsf{#1}}}}

\newcommand{\system}{SEFL\xspace}

\newcommand{\AS}{\ensuremath{\mathsf{AS}}\xspace}
\newcommand{\CSP}{\ensuremath{\mathsf{CSP}}\xspace}

\newcommand{\ignore}[1]{}

\newcommand{\pk}{\ensuremath{\class{pk}}\xspace}
\newcommand{\sk}{\ensuremath{\class{sk}}\xspace}

\newcommand{\M}{\ensuremath{\mathcal{M}}}

\newcommand{\AHE}{\ensuremath{\class{AHE}}\xspace}

\newcommand{\enc}{\ensuremath{\class{Enc}}\xspace}
\newcommand{\Enc}{\ensuremath{\class{Enc}}\xspace}
\newcommand{\Gen}{\ensuremath{\class{Gen}}\xspace}

\newcommand{\Dec}{\ensuremath{\class{Dec}}\xspace}

\usepackage[]{emnlp2021}
\usepackage{graphicx}
\usepackage{times}
\usepackage{algorithmicx}
\usepackage{algpseudocode}
\usepackage{latexsym}
\usepackage{amsfonts}
\usepackage{amsmath}
\newtheorem{proof}{Proof}

\usepackage[T1]{fontenc}

\usepackage[utf8]{inputenc}

\usepackage{microtype}

\title{A Secure and Efficient Federated Learning Framework for NLP}

\author{Jieren Deng$^{1}$\textsuperscript{\textsection}, Chenghong Wang$^{2}$\textsuperscript{\textsection}, Xianrui Meng$^3$, Yijue Wang$^1$, Ji Li$^4$, Sheng Lin$^5$\\ \textbf{Shuo Han$^6$, Fei Miao$^1$, Sanguthevar Rajasekaran$^1$, Caiwen Ding$^1$} \\
        $^1$University of Connecticut, $^2$Duke University, $^3$Facebook, 
        $^4$Microsoft\\
        $^5$Northeastern University,
        $^6$University of Illinois at Chicago\\
        \small\texttt{\{jieren.deng,yijue.wang,fei.miao,sanguthevar.rajasekaran,caiwen.ding\}@uconn.edu}\\
        \small\texttt{chenghong.wang552@duke.edu,\{xianruimeng,changzhouliji\}@gmail.com}\\ \small\texttt{lin.sheng@northeastern.edu, hanshuo@uic.edu}}

\begin{document}
\maketitle

\begin{abstract}
 In this work, we consider the problem of designing secure and efficient federated learning (FL) frameworks. Existing solutions either involve a trusted aggregator or require heavyweight cryptographic primitives, which degrades performance significantly. Moreover, many existing secure FL designs work only under the restrictive assumption that none of the clients can be dropped out from the training protocol. To tackle these problems, we propose \system, a secure and efficient FL framework that (1)~eliminates the need for the trusted entities; (2)~achieves similar and even better model accuracy compared with existing FL designs; (3)~is resilient to client dropouts. 
  Through extensive experimental studies on natural language processing (NLP) tasks, we demonstrate that the \system achieves comparable accuracy compared to existing FL solutions, and the proposed pruning technique can improve runtime performance up to 13.7$\times$.
\end{abstract}
\begingroup\renewcommand\thefootnote{\textsection}
\footnotetext{Equal contribution, alphabetical order}
\section{Introduction}
Deep Neural Networks have played a significant role in advancing many applications~\cite{9474235,10.1145/3123939.3124552}. The field of Natural Language Processing (NLP) leverages Recurrent Neural Networks (RNNs) and Transformers to achieve outstanding performance on many tasks. 
The Transformer was first introduced in \cite{vaswani2017attention} using a self-attention mechanism and it achieved prominent performance
in various NLP tasks. 
The benefits of RNNs and Transformers in NLP are well-publicized, but the various privacy and security problems still pose challenges to the utilization of these models  by data owners, especially users with sensitive data such as location, health, and financial datasets.  
{\em Federated Learning} (FL)~\cite{google-fl} empowers different data owners (e.g., organizations or edge devices) to collaboratively train a model without sharing their own data, thus allowing them to address key issues like data privacy. Although data exchanged in FL consists of less information of the user's raw data~\cite{bonawitz2019towards}, one might still be concerned about how much information remains. Recent research has shown that attackers can still infer sensitive information about the training data, or even reconstruct the it solely from publicly shared model parameters~\cite{zhu2019deep}.

Although a series of works~\cite{bonawitz2017practical,truex2019hybrid,conf/iclr/PapernotSMRTE18,wu2021novel,lin2020esmfl} have been proposed to protect FL protocols from leaking sensitive information~\cite{ijcai2021-432,deng2021tag,wang2020sapag}. They either have to involve a trusted third party (centralized aggregator), or do not tolerate client dropouts. Therefore, the data owners either need to blindly trust the centralized aggregator or must be online all the time during the training period, which makes the entire design less practical. To address the aforementioned issues, in this work, we develop a secure and efficient FL framework, \system. It employs two non-colluding servers, i.e., \textit{Aggregation Server} (\AS) and \textit{Cryptography Service Provider} (\CSP). \AS collects the encrypted local updates from clients, and securely aggregates them, while \CSP manages the cryptography primitives, i.e. the decryption key. The overarching goal of this framework is to support accurate and efficient RNN and Transformer training while preserving the privacy of training data against the untrusted servers. In other word, any servers' knowledge about any single training data should be bounded by differential privacy~\cite{10.5555/1791834.1791836}.


 \begin{figure*}
\centering
  \centering
  \includegraphics[width=0.9\linewidth]{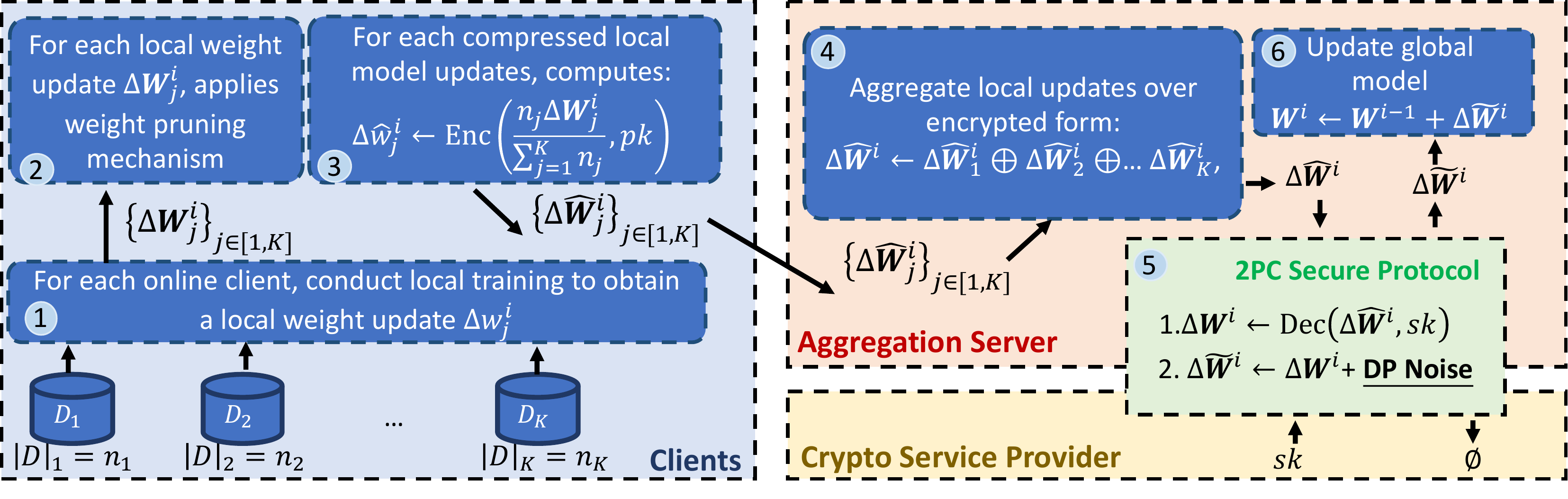}
  \captionof{figure}{\system workflow}
  \label{fig:overview}
\end{figure*}

\ignore{
To securely protect local updates from clients, we adopt \textit{additively homomorphic encryption} (AHE). In comparison to Multiparty Computation (MPC) solutions~\cite{bonawitz2017practical}, not only our design can make it easy for users to protect their updates using encryption, but it also makes the user enrollment phase much more simpler and flexible. Moreover, this results in a better solution when handling dropped out clients. We show that our framework is more robust to disconnected users. 
Our protocol can handle an arbitrary number of dropped out clients as long as the number of remaining clients can produce a useful model.
Cryptographic primitives can help us to provide stronger security guarantees. However, in practice, they often come at very high computation and communication overhead. Thus, to design a large-scale and provably secure FL framework is considerably challenging, especially given that the current ML model gets bigger and more complex.
For large ML  models, there exist (i) a vast amount of data communication between the edge devices and cloud, and (ii) limited resources (e.g., computation, memory size) on edge devices. Adding additional cryptographic operations in an FL framework could potentially prohibit the popularity and the adoption of large-scale edge devices such as mobile or IoT.
}
\eat{Our new framework \system significantly improves the security, efficiency, and utility of the existing FL works. Our contributions are summarized as follows.}
   Our contributions are summarized as follows:
 (1) We present 
    a novel secure FL framework that eliminates the need for trusted aggregators.
    (2) \system is more resilient to clients dropping out than previous works. \system is able to produce a correct global model even 75\% of clients are dropped out from the training protocol.  
   (3) To improve the training performance, we integrate the Hankel-matrix based local update/weight pruning method with \system  to simultaneously  reduce  the  volume  of local  update and  weight  storage. The reduction in space, computational, and communication complexity are significant, from O($l^2$) to O($2l-1$) for weight/update representation, where $l$ is the block size. 
With extensive experiments, we show that \system achieves comparable or even better accuracy than existing secure FL solutions over complex RNN and Transformer models, and the proposed pruning scheme improves \system's performance up to 13.7$\times$.


\eat{The rest of this paper is organized as follows. We discuss the background and technical details in Section~\ref{sec:background} and~\ref{sec:sysdesc}, respectively. In Section~\ref{sec:compress} we introduce the block Hankel-matrix based performance optimization to our private FL design for high scalability. We include the security analysis sketch in Section~\ref{sec:secsk}. We provide comprehensive experimental evaluations in Section~\ref{sec:experiments}. Finally, we give an overview of related work and concluding remarks in Section~\ref{sec:relatedworks} and Section~\ref{sec:conslusion}.}
\section{Background}\label{sec:background}
\noindent{\bf Differential privacy.} Let $\epsilon, \delta > 0$ be privacy parameters, a randomized mechanism $\mathcal{M}$ satisfies $\epsilon, \delta$-differential privacy ($\epsilon, \delta$-DP) if and only if for any two adjacent datasets $D$ and $D'$ (differ by addition or removal of one data), for any possible output $S$, the following holds: 
\[
\Pr \big[\M(D) \in S\big] \leq e^{\epsilon}Pr\big[\M(D') \in S\big] + \delta
\]
The Gaussian Mechanism (GM)~\cite{dwork2014algorithmic} achieves differential privacy by approximating a deterministic real-valued function $f$ with an additive noise that is proportional to the function's sensitivity $S_f$, where $S_f = \max_{D, D'}|f(D) - f(D')|$. A GM is written as $\M(D) = f(D) + \mathcal{N}(0, \sigma^2S^2_f)$, where $\mathcal{N}$ denotes a normal distribution, and $\sigma$ is the noise scale.

\noindent{\bf Additively homomorphic encryption ($\AHE$).} $\AHE$ is a semantic secure public-key encryption scheme~\cite{Peter2012AdditivelyHE}, with three algorithms $\Gen$, $\Enc$ and $\Dec$, where $\Gen$ generates they public and secret key pairs $(\pk, \sk)$, $\Enc$ encrypts a message with $\pk$ and $\Dec$ decrypts a ciphertext with secret key $\sk$. In addition $\AHE$ provides a homomorphic addition operator $\oplus$, such that  $\Dec(\Enc(m_1, \pk) \oplus \Enc(m_2, \pk) ...\oplus \Enc( m_k, \pk), \sk) = m_1 + \dots + m_k$. 

\noindent{\bf Two party secure computation (2PC).} 2PC allows two parties with private inputs $x_1$ and $x_2$ to jointly compute a given function $f$. Both parties learn nothing beyond the output of $f$. A typical 2PC design is the garbled circuit (GC) \cite{Yao}.


\section{\system Explained}

\subsection{Workflow}

We design \system framework based on the two non-colluding (untrusted) server setting, where an aggregation server (\AS) aggregates the encrypted local model updates and another sever (\CSP) manages the cryptography primitives (i.e. the decryption key). To ensure the privacy, we require that any server's knowledge about any single training data is bounded by some differential privacy. Figure~\ref{fig:overview} illustrates an overview of \system. 

Initially, \CSP generates the key pairs $(pk, sk)$, stores the secret key $sk$ locally, and broadcasts the public key $pk$ to all other entities (\AS and all clients). In our design, \CSP is tasked to manage the cryptography primitives (i.e. the $\sk$), thus \CSP is the only entity that can decrypt the encrypted messages under the secret key $\sk$. In the meantime, we assume that all entities will agree on a same initial model ${\bf W}^0$.

Each training iteration, i.e. $i^{th}$ training round, starts with all clients conduct local training with their respective private data $D_j$ with a data size $n_j$ then obtain the local model update $\Delta {\bf W}^{i}_j$. Then, each client prunes the obtained model updates using weight pruning techniques and encrypts the compressed update by computing $\Delta \hat{{\bf W}}^{i}_j \gets\enc(\frac{n_j\Delta {\bf W}^i_j}{\sum_{j=1}^{K}n_j}, pk)$. Clients then submit the encrypted and compressed updates to the \AS.

On the server side, \AS homomorphically adds all encrypted (pruned) local updates over encrypted form and then obtains $\Delta \hat{{\bf W}}^i\gets \Delta \hat{{\bf W}}^i_1 \oplus \Delta \hat{{\bf W}}^i_2 \oplus ...$. Knowing that $\Delta \hat{{\bf W}}^i$ is equal to the encryption of the weighted average of all pruned local updates, that is  $\Delta \hat{{\bf W}}^i = \enc(\sum_{j=1}^{K} \frac{n_j\Delta {\bf W}^{i}_j}{\sum_{j=1}^K n_j} , pk)$.  To decrypt the aggregated global update, \AS has to collaborate with \CSP, as \CSP is the only entity that manages the decryption key. Moreover, sending $\Delta \hat{{\bf W}}^i$ directly to \CSP for decryption will result in the exact value of $\Delta {\bf W}^i$ being exposed to \CSP, which violates the privacy guarantee. One possible approach is to have \AS homomorphically add some random noise to $\Delta \hat{{\bf W}}^i$ and send the distorted global update to \CSP for decryption. After receiving the result of decryption, \AS removes the noise to obtain the true answer. This prevents \CSP from knowing the true value of the global model update, however \AS will know this value, which is also a privacy violation. To ensure none of the two servers can learn the exact global updates,  in our design, \AS first sends a distorted $\Delta \hat{{\bf W}}^i$ (with some random mask) to \CSP, followed by \CSP decrypts the distorted global update. Then, the two servers jointly evaluate a secure 2PC where \AS inputs the random mask and \CSP inputs the decrypted global update. $\Delta \hat{{\bf W}}^i$ is then recovered inside the secure 2PC protocol. Next, each server independently samples a DP noise and provides it as input to the secure 2PC. These DP noises are then added to the recovered $\Delta \hat{{\bf W}}^i$ inside the 2PC protocol. Finally, the protocol returns the global update distorted by DP noise to \AS, with which \AS updates the global model, ${\bf W}^i\gets {\bf W}^{i-1} + \Delta \tilde{\bf W}^i$. Note that, the choice of DP noise is quite flexible, and by default, \system uses Gaussian noise to distort the global update. 

\system repeats the training phases until it reaches the maximum training round $T$ or the model is converging. Note that, it is not necessary for \AS to have all local updates from clients, according to our evaluation results,
\system is able to train an accurate model when only 10\% of clients contribute their local updates. Therefore, in practice, one can set an aggregation threshold, say $L$, which means that \AS can start aggregating local updates as long as it receives more than $L$ updates.



\eat{A comprehensive description of \system's workflow is described in Algorithm~\ref{algo:sf}.}

\eat{
\begin{figure*}[t]
\centering
\includegraphics[width = 0.8\textwidth]{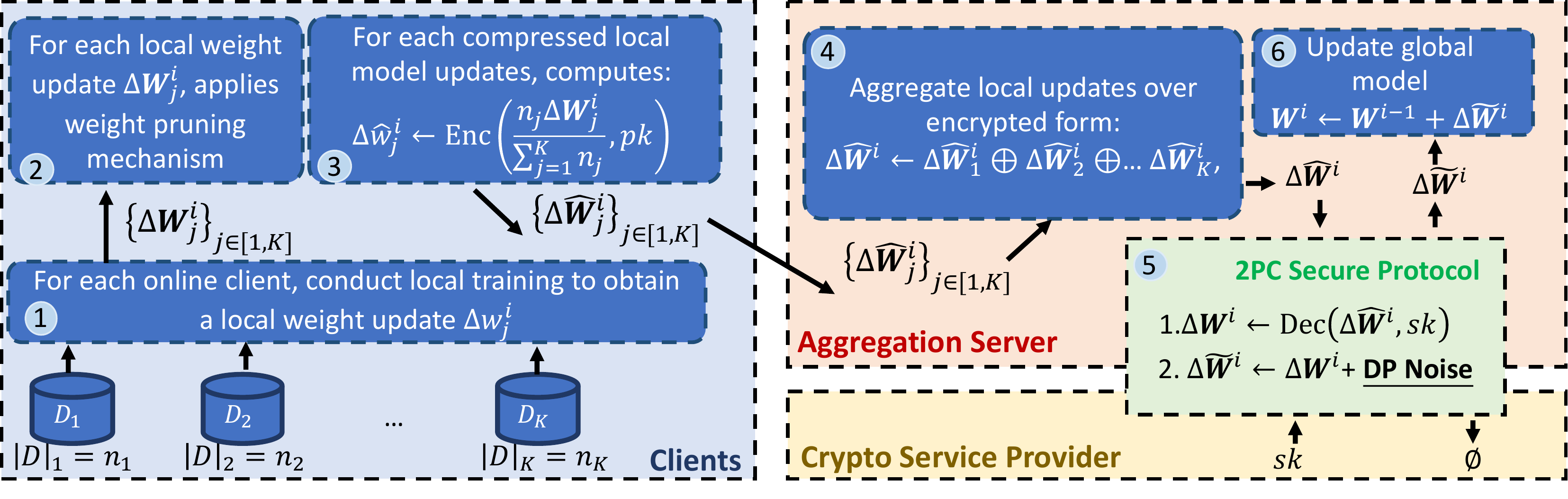}\vspace{-2mm}
\caption{\system Framework Overview.
}
\label{fig:overview}
\end{figure*}
}

\begin{figure}
  \centering
  \includegraphics[width=0.9\linewidth]{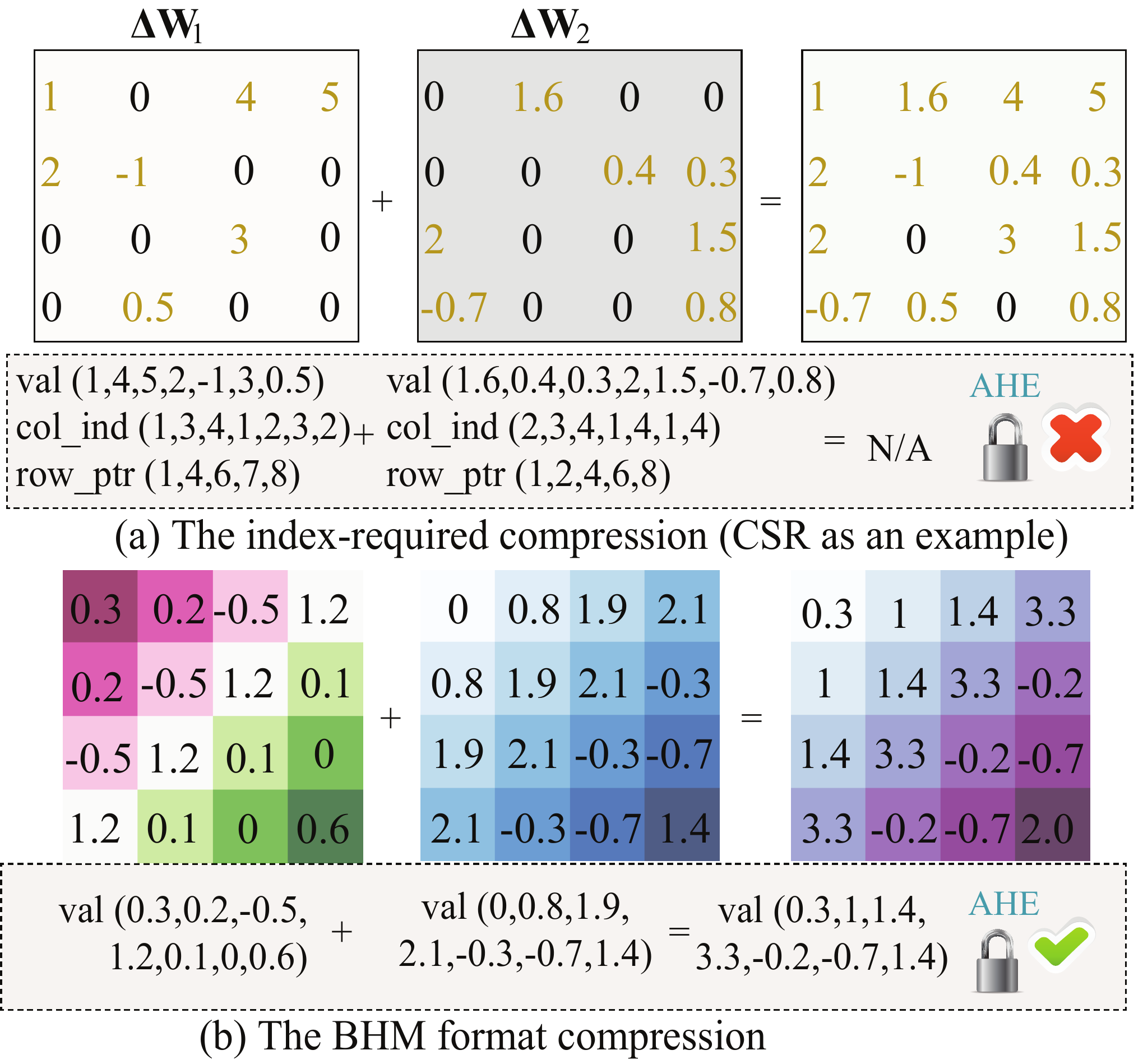}
  \captionof{figure}{CSR format vs. BHM format}
  \label{fig:Crypto_demo}
\end{figure}

\subsection{Block-Hankel Matrix-based Pruning}
Cryptographic primitives can help to provide stronger security guarantees. However, in practice, they often come at high computation and communication overhead. Adding additional cryptographic operations in an FL framework could potentially prohibit the popularity and the adoption of resource-constrained edge devices such as mobile or IoT devices with limited resources (e.g., computation, memory size). Therefore, to be compatible with resource-constrained edge devices on federated learning, we aim to minimize the number of cryptographic operations required during training while maintaining the accuracy of FL. To achieve this, we develop an efficient method to train a large neural network by simultaneously reducing the volume of local updates and weight storage. 
we design an efficient method to train a large NLP model with reduced volume of local updates, to reduce the number of required cryptographic operations.

\noindent\textbf{Pitfall of sparsity format in AHE.} 
Typical weight pruning approaches require to store the indices of nonzero entries~\cite{Gurevin2021Enabling,gui2019model,wen2016learning,ren2020darb,ma2020pconv}.
However, 
the different position of nonzero values from all clients 
can lead to significant inefficiency for the subsequent model update aggregation.
\eat{
\begin{figure}[t]
\centering
\includegraphics[width = 0.43\textwidth]{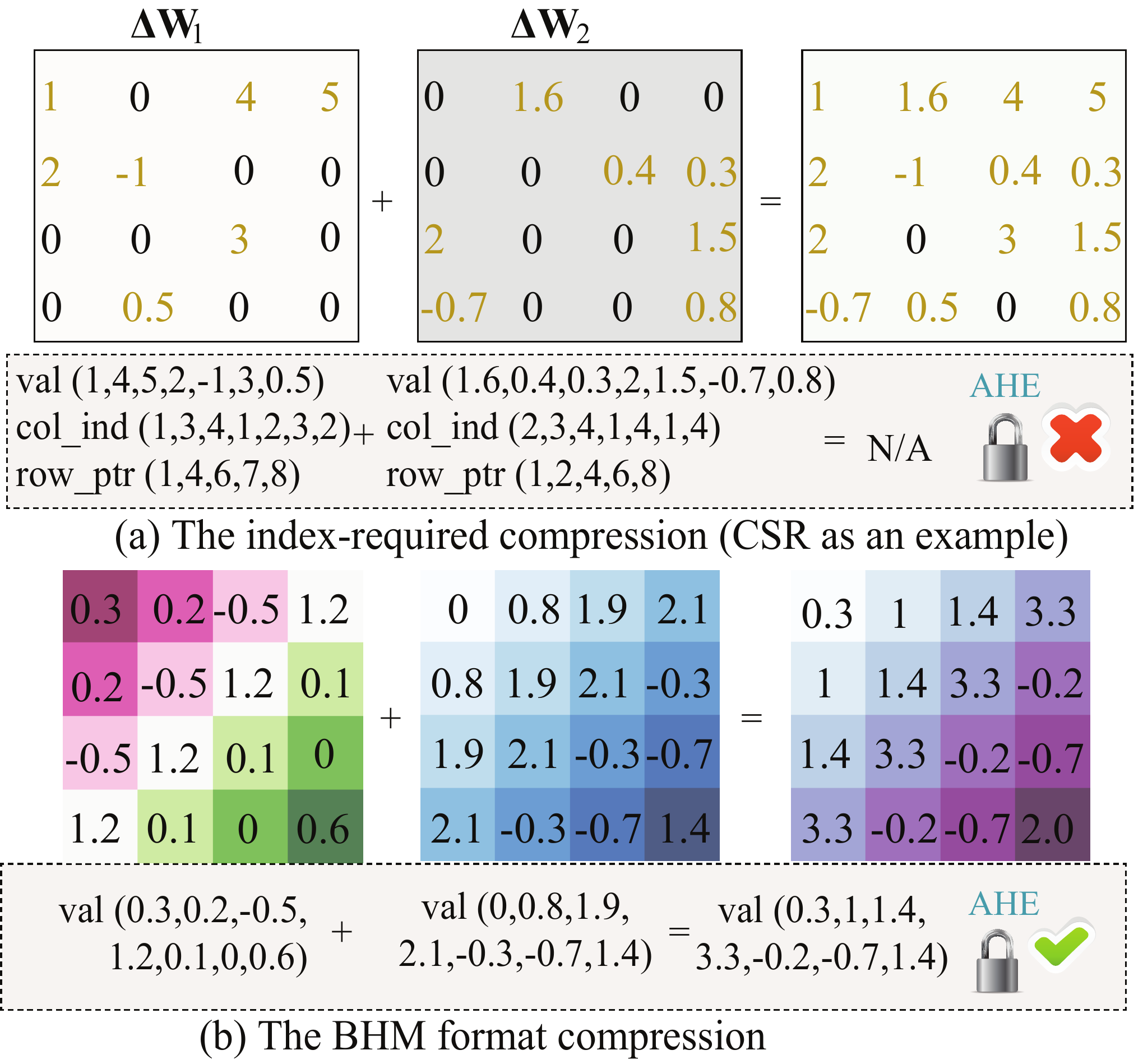}
\caption{(a) The CSR format and (b) the proposed BHM format sparsity in AHE.}
\label{fig:Crypto_demo}
\end{figure}
}
As shown in Fig.~\ref{fig:Crypto_demo} (a), assume \AS aggreates two local updates with the same sparsity from $\mathcal{C}_1$ and $\mathcal{C}_2$. We apply compressed sparse row (CSR) format, to represent the updates ($\Delta {\bf{W}}_1$ and $\Delta {\bf{W}}_2$ in Fig.~\ref{fig:Crypto_demo} (a)), where the non-zero elements of $\Delta {\bf{W}}_1$ and $\Delta {\bf{W}}_2$
are not located in the same position. As the AHE-based update aggregation process is a black-box homomorphic addition operation, we can not reconstruct the original sparse matrix from CSR since indices are encrypted, therefore we can not correctly produce the aggregated update. 

 \noindent\textbf{Crypto-friendly Block-Hankel matrix based pruning.} 
 We divide the local update into multiple modules with identical shape. Within each module, a special format of structure matrix is applied to approximate the original matrix without indices. 
 In our framework, we investigate the use of blocks of Hankel matrix (BHM) to approximate blocks of local update. As shown in Fig.~\ref{fig:Crypto_demo} (b), we can perform aggregation based on the encrypted $val$ vectors since the positions of the sequence vectors are identical. In addition, the resultant global model will have the same size,
therefore downloading  and  uploading communication is symmetric and balanced.

\eat{
\begin{equation}
\mathbf{\bf{g}}_{mn}
=
\begin{bmatrix}
 g_{0}      & g_{1}      & \cdots & g_{l-2}  & g_{l-1}      \\
 g_{1}      & g_{2}      & \cdots  & g_{l-1} & g_{l}      \\
 \vdots & \vdots & \ddots & \vdots & \vdots\\
 g_{l-2}      & g_{l-1}      & \cdots & g_{2l-3} & g_{2l-3}      \\
 g_{l-1}      & g_{l}      & \cdots & g_{21-3}  & g_{2l-2}      \\
\end{bmatrix},
\end{equation}

We use $l$ to represent the row/column size of each Hankel matrix (or block size). 
We use the compression of local update as a demonstration. 
The corresponding weight/models has the same size and compression form. 
Each sub-matrix ${\bf g}_{ij}\in\mathbb{R}^{l \times l}$ is defined by the Hankel operator ${{H_d}}\in\mathbb{R}^{2l-1}\rightarrow \mathbb{R}^{l \times l}$, and can be represented by the sequence vector ${\bf g}_{ij}\in\mathbb{R}^{2l-1}$. }

\eat{
Our main goal is to find a pruned network $\hat{g}(x)$ which is competitive with $f(x)$,i,e:
\begin{equation}
 \underset{x \in \chi}{\text{sup}} {\left \| f(x) - \hat{g}(x) \right \|} \leq \epsilon   
\end{equation} 
By using our proposed pruning method, the result for finding $\hat{g}(x)$ is fallowing. We prove that from sufficiently logarithmically over-parameterized neural network with random initialized weights there is a subnetwork can match the same performance with high probability.
}

In what follows, we discuss the convergence analysis for pruned sub-networks.

{\bf Theorem 1.} {\it For every network $f$ with depth $l$ and $\forall~ i \in \{1,2,\dots,n\}$. Consider g is a randomly initialized neural network with 2n layers, and width $poly(d,n,m)$, where $d$ is input size, n is number of layers in $f$, m is the maximum number of neurons in a layer.  The weight initialization distribution belongs to Uniform distribution in range [-1,1]. Then with probability at least $1-\beta$  there is a weight-pruned subnetwork $\hat{g}$ such that:} 
\begin{equation}
 \underset{x \in \chi, \left \| W \right \| \leq 1}{\text{sup}} {\left \| f(x) - \hat{g}(x) \right \|} \leq \alpha   
\end{equation}
\eat{
{\bf Theorem 2.~\cite{lueker1998exponentially}} {\it Let $W^*_1,...,W^*_n$ as $n$ i.i.d. random variables drawn from the Uniform distribution over [-1,1], where $n \geq Clog\frac{2}{\delta}$ ,where $\delta \leq min\{1,\epsilon\}$. Then, with probability at least 1-$\delta$, we have \\
\begin{equation}
\begin{aligned}
    &\exists S \subset \{1,2,...,n\}, \forall W \in [-0.5,0.5],\\
    &s.t \left | W-\sum_{i \in S}W^*_i \right | \leq \epsilon 
\end{aligned}
\end{equation}
}
}
\vspace{-2mm}
\begin{proof}We start with analysis over simple ReLU networks, where $f(x) = w \cdot x$, $g(x) = \mathbf{u}\sigma (\mathbf{w}^g x)$. Since $\sigma$ is a ReLU activation function, thus $w =\sigma (w) - \sigma(-w)$ and such that  $x^*  \mapsto \sigma \left ( wx \right ) = \sigma\left ( \sigma(wx)  - \sigma(-wx) \right)$.
On the other hand, this neuron can be present as: $x^*  \mapsto  \mathbf{u} \sigma \left ( \mathbf{p}\odot  \mathbf{w}^g x\right )$. Let $\mathbf{w^+}=\max\{\mathbf{0}, \mathbf{w}\}$, $\mathbf{w^-}=\min\{\mathbf{0},\mathbf{w}\}$, $\mathbf{w^+}+\mathbf{w^-}=\mathbf{w}^g$. Then
\begin{equation}
    x^*  \mapsto  \mathbf{u} \sigma\left( \sigma \left ( \mathbf{p}\odot  \mathbf{w^+}x\right )-\sigma \left ( \mathbf{p}\odot  \mathbf{-w^-}x\right )  \right ) 
\end{equation}
Base on~\cite{lueker1998exponentially}, when $n \geq C\log{\frac{4}{\alpha}}$, $\forall w^f \in [0,1]$, there exist a pattern of $\mathbf{w}$ and $ p \in \{0,1\}^n$:
\begin{equation}
    \textup{Pr} \left[ \left | w^f- \mathbf{u} \sigma( \mathbf{p} \odot \mathbf{w^+}) \right | < \frac{\alpha}{2}\right]\geq1-\frac{\beta}{2}
    \label{wplus}
\end{equation}
By symmetric, Eq.~\ref{wplus} holds for $\mathbf{w^-}$ as well. Therefore, we obtains $\text{sup} \left | w^fx- \mathbf{u} \sigma( \mathbf{p} \odot \mathbf{w}x) \right | \leq \alpha$. To extend it to a single network layer, we computes 
\begin{equation}
\begin{split}
&\text{sup}
\left | \mathbf{W}^f\mathbf{x}- \mathbf{u} \sigma( \mathbf{p} \odot \mathbf{W}^g\mathbf{x}) \right | \\
&\leq \sum_{j=1}^{k}\sum_{i=1}^{m} \text{sup} \left | w_{j,i}^fx_i- \mathbf{u}_i \sigma( \mathbf{p}_{j,i} \odot \mathbf{w}_{j,i}x_i) \right | \leq \alpha\\ 
\end{split}
\label{conv_layer}
\end{equation}
\vspace{-1mm}
We now provide the general case analysis. With probability over $1-\beta$, we obtains:
\begin{equation}
\small\begin{split}
& \left \| f(x)-\hat g(x) \right \| \\ 
= & \left \| \mathbf{W}_n \mathbf{x}_n - \mathbf{P}_{2n} \odot \mathbf{W}^g_{2n}\mathbf{x}^g_n \sigma(\mathbf{P}_{2n-1} \odot \mathbf{x}^g_{2n-1}) \right \|  \\
\leq &  \alpha/2+ \alpha/2 = \alpha\\
\end{split}
\end{equation}
\end{proof}

\eat{
\textbf{Case 3. Analysis on a Neuron}\\
\begin{equation}
\begin{split}
&\text{sup}
\left | \mathbf{w}^f\mathbf{x}- \mathbf{u} \sigma( \mathbf{p} \odot \mathbf{wx}) \right | \\
&\leq \text{sup} \left | \sum_{i=1}^{m}\left (  w_i^fx_i- \mathbf{u}_i \sigma( \mathbf{p}_i \odot \mathbf{w}_ix_i)\right ) \right | \\ 
&\leq \text{sup}\sum_{i=1}^{m} \left | w_i^fx_i- \mathbf{u}_i \sigma( \mathbf{p}_i \odot \mathbf{w}_ix_i) \right | \\ 
&\leq \sum_{i=1}^{m} \text{sup} \left | w_i^fx_i- \mathbf{u}_i \sigma( \mathbf{p}_i \odot \mathbf{w}_ix_i) \right | \\ 
&\leq  m \cdot  \frac{\epsilon}{m}\\
&\leq \epsilon
\end{split}
\label{conv_neuron}
\end{equation}

\textbf{Case 4. Analysis on a single network Layer}\\
\begin{equation}
\begin{split}
&\text{sup}
\left | \mathbf{W}^f\mathbf{x}- \mathbf{u} \sigma( \mathbf{p} \odot \mathbf{W}^g\mathbf{x}) \right | \\
&\leq \text{sup} \left | \sum_{j=1}^{k}\sum_{i=1}^{m}\left (  w_{j,i}^fx_i- \mathbf{u}_i \sigma( \mathbf{p}_{j,i} \odot \mathbf{w}_{j,i}x_i)\right ) \right | \\ 
&\leq \text{sup}\sum_{j=1}^{k}\sum_{i=1}^{m} \left | w_{j,i}^fx_i- \mathbf{u}_i \sigma( \mathbf{p}_{j,i} \odot \mathbf{w}_{j,i}x_i) \right | \\ 
&\leq \sum_{j=1}^{k}\sum_{i=1}^{m} \text{sup} \left | w_{j,i}^fx_i- \mathbf{u}_i \sigma( \mathbf{p}_{j,i} \odot \mathbf{w}_{j,i}x_i) \right | \\ 
&\leq  k \cdot m  \cdot \frac{\epsilon}{mk}\\
&\leq \epsilon
\end{split}
\label{conv_layer}
\end{equation}

\textbf{Case 5. Putting it all together}\\
We now provide the general case analysis, where $f(x)$ is defined as Eq.\ref{def_f},  $g(x)$ is defined as Eq.\ref{def_g}. With probability over $1-\epsilon$, we can conclude that:
\begin{equation}
\begin{split}
& \left \| f(x)-\hat g(x) \right \| \\ 
& =\left \| \mathbf{W}_n \mathbf{x}_n - \mathbf{P}_{2n} \odot \mathbf{W}^g_{2n}\mathbf{x}^g_n \sigma(\mathbf{P}_{2n-1} \odot \mathbf{x}^g_{2n-1}) \right \|  \\
& \leq  \left \| \mathbf{W}_n\mathbf{x}_n -  \mathbf{W}_n\mathbf{x}_n^g \right \| + \\ 
& \left \| \mathbf{W}_n\mathbf{x}_n^g -  \mathbf{P}_{2n} \odot \mathbf{W}^g_{2n}\mathbf{x}^g_n \sigma(\mathbf{P}_{2n-1} \odot \mathbf{x}^g_{2n-1}) \right \|  \\
& \leq \left \| \mathbf{x}_n-\mathbf{x}_n ^g \right \| + \\
& \left \| \mathbf{W}_n\mathbf{x}_n^g -  \mathbf{P}_{2n} \odot \mathbf{W}^g_{2n}\mathbf{x}^g_n \sigma(\mathbf{P}_{2n-1} \odot \mathbf{x}^g_{2n-1}) \right \|  \\
& \leq \epsilon/2+ \epsilon/2\\
& \leq \epsilon
\end{split}
\end{equation}
}

\noindent\textbf{Putting it all together.} 
Our objective is to compress the weights and updates using the BHM formats. Thus we minimize the loss function subject to constraints of  BHM. More specifically, we set constraints as \smash{${\bf{S}}_{i}^{(t)}= \{ {\bf{W}}_{i}^{(t)}\mid  {\bf{W}}_{i}^{(t)}\in \text{BHM} \}$}.
The \emph{backward propagation} process of the training phase can also be implemented using the BHM format, since pruning based on the block Hankel matrix has the same ``effectiveness'' as unpruned DNNs, as shown in~\cite{zhao2017theoretical}. 
 

Compared to other index-required pruning methods, the BHM pruning has the following advantages. First,  it always guarantees the strong structure of the trained network, thereby avoiding the storage space, computation, and communication time overhead incurred by the complicated indexing process. 
Second, 
during training, the BHM-based approach directly trains weight matrices in the BHM format by updating only one vector for each block (i.e., 
$2l-1$ vs. $l^2$). 
Third, the reduction in space, computational, and communication complexity by using BHM are significant. The weight tensor \smash{$ {\bf{W}}_{i}^{(t)}$} and updates \smash{$\Delta {\bf{W}}^{(t)}_{i}$} have the storage complexity and communication complexity reduced from O($l^2$) to or O($2l-1$).

\section{Experiments}
We implement the \system system using PyTorch 1.4.0, CUDA 10.1. All experiments are performed on the AWS EC2 cloud instance with a 2.30GHz Intel Xeon Gold 5218 Salable Processors and 8 NVIDIA Quadro RTX 6000 GPUs.
We evaluate \system by conducting experiments using LSTM and Transformer on WikiText-2~\cite{merity2016pointer} dataset. The LSTM model is adopted from~\cite{HochSchm97}.
The Transformer model~\cite{vaswani2017attention} contains two layers with an embedding dimension of 200, two attention heads, and 200 hidden units. We use perplexity to measure the quality of the predicted data for both Transformer and LSTM.
\begin{figure}[t]
\centering
\includegraphics[width = 0.5\textwidth]{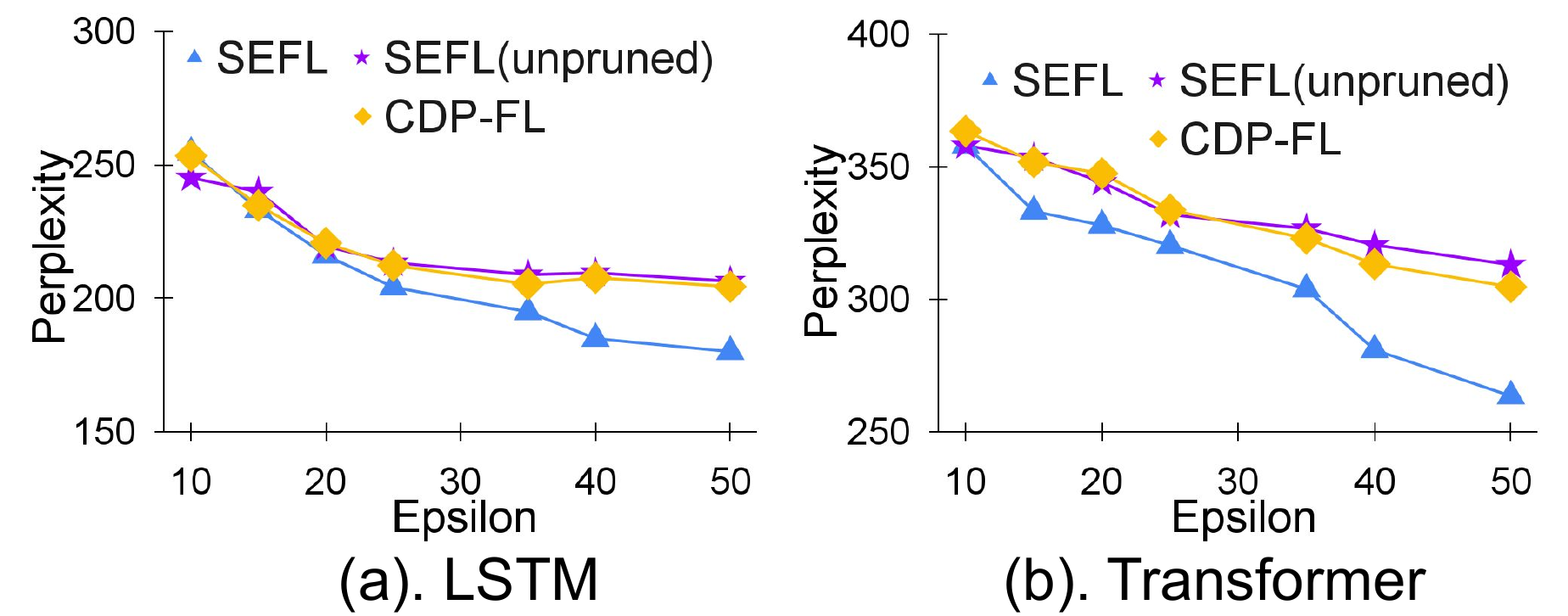}
\caption{Comparison of the impact of different federated learning approaches on accuracy.}


\label{fig:epsilon}
\end{figure}
\begin{figure}[t]
\centering
\includegraphics[width = 0.5\textwidth]{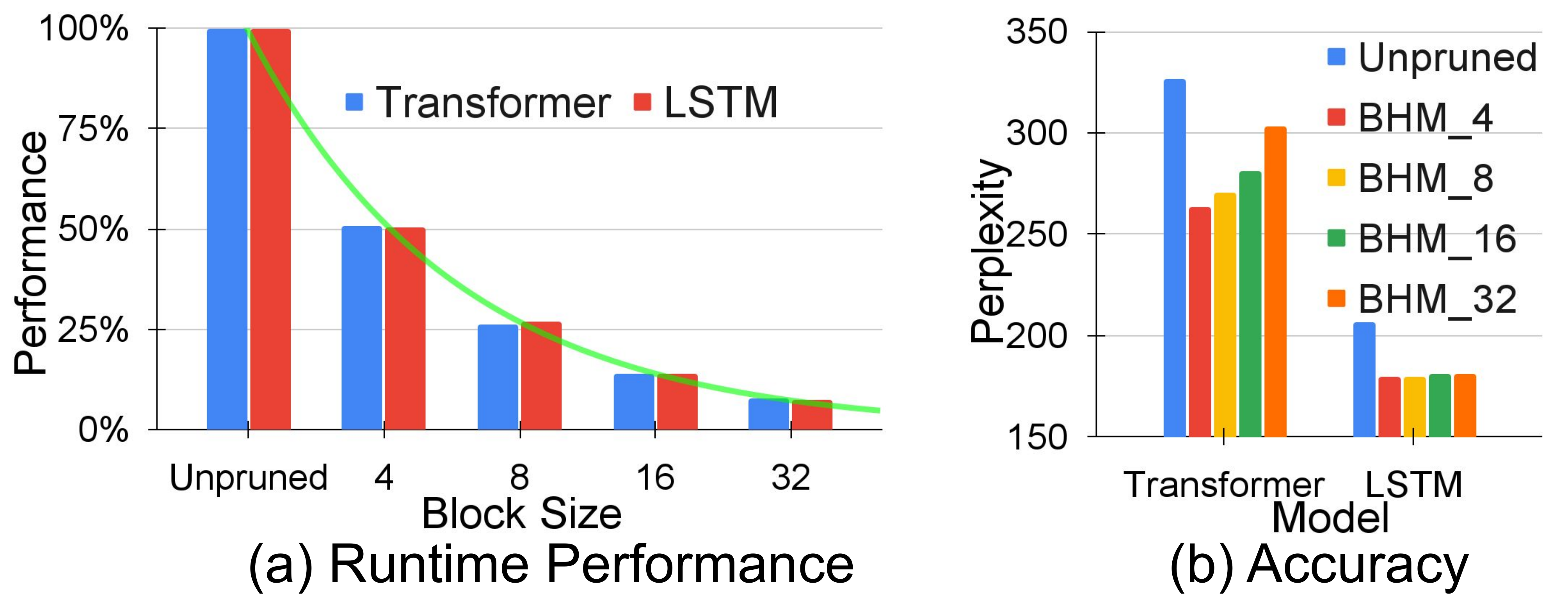}
\caption{Comparison of the impact of different Block Size on accuracy and performance}

\label{fig:block_size}
\end{figure}
\subsection{Result Analysis}
\noindent\textbf{Comparisons with existing private FL.} 
In Figure~\ref{fig:epsilon}, we compare SEFL with the state-of-art private FL design, CDP-FL~\cite{geyer2017differentially}.
First, the unpruned \system achieves similar accuracy compared to the CDP-FL due to the fact that the logical training process of these two approaches are similar. Both methods first utilize FedAvg~\cite{mcmahan2017communicationefficient} to obtain a global aggregation model and then distort it with DP noise. The optimized (with pruning technique) \system improves the accuracy by up to 11\% and 15\% over the CDP-FL method, with LSTM and Transformer model, respectively. One possible explanation is that pruning reduces the DP noises added to the aggregation model. Since, to distort the model, one should inject Gaussian noise to each model element independently. Therefore, the smaller the size of the model, the fewer times the Gaussian noise is injected. 

\noindent\textbf{Evaluating optimization.}
We compare the \system (with pruning optimization) with unoptimized \system in Figure~\ref{fig:block_size}. We report the average elapse time in seconds over 10 replicated runs as the runtime performance. We report the accuracy and runtime performance for unpruned \system and \system with BHM block size from 4 to 32. Note that the larger the block size, the smaller the compressed model will be. 
\system achieves a performance improvement of up to 13.7$\times$ over the unpruned \system under both LSTM and Transformer models. Additionally, \system shows better accuracy (smaller perplexity) compared to the unpruned \system in almost all test groups. For best cases, \system achieves 19.3\%, and 12.8\% accuracy improvement,respectively, in contrast to the unoptimized \system implementation. 
BHM-based pruning optimization not only brings significant performance improvements, but also optimizes the accuracy guarantees.


\begin{table}[]
\resizebox{\columnwidth}{!}{%
\begin{tabular}{l|llll}
\hline
\multicolumn{1}{c|}{\begin{tabular}[c]{@{}c@{}}Dropout\\ Rate\end{tabular}} &
  \begin{tabular}[c]{@{}l@{}}BHM\\ (75\%)\end{tabular} &
  \begin{tabular}[c]{@{}l@{}}BHM\\ (50\%)\end{tabular} &
  \begin{tabular}[c]{@{}l@{}}BHM\\ (25\%)\end{tabular} &
  \begin{tabular}[c]{@{}l@{}}BHM\\ (0\%)\end{tabular} \\ \hline
LSTM &
  194.82 &
  187.64 &
  178.45 &
  177.41 \\ \hline
Transformer &
  310.12 &
  301.04 &
  279.81 &
  263.78 \\ \hline
\end{tabular}}
\caption{Comparison of the impact of dropout rate.}
\label{tab:drop_out}
\end{table}

\noindent\textbf{\system with clients dropout.} We evaluate whether \system is able to handle clients dropouts in Table~\ref{tab:drop_out}, where we report accuracy when  25\%, 50\%, and 75\% of clients are dropped out from the protocol. Shown in Table~\ref{tab:drop_out}, when the dropout rate is relatively small, i.e., 25\%, \system achieves almost the same accuracy guarantee as in the no-dropout case (0.5\% and 5\% accuracy degradation for LSTM and Transformer, respectively). 
Even when majority of clients are drooped out, i.e. 75\% drop rate, \system still produces accurate models with only 17.41 and 46.34 higher perplexity. In summary, \system can handle a large number of client dropouts with relatively small degradation in accuracy. This result shows that our proposed approach is applicable to practical scenarios.
\section{Conclusion}
In this paper, we introduced a new secure and efficient FL framework, \system, that (i) eliminates the need for the trusted entities, (ii) achieves similar model accuracy compared with existing FL approaches, and (iii) is resilient to client dropouts. We also proposed optimizations that mitigate the high computation and communication overhead caused by cryptographic primitives. This is achieved by applying a local weight pruning technique based on the block Hankel-matrix. Through extensive experimental studies on NLP tasks, we demonstrate that the \system achieves comparable accuracy compared to existing FL solutions, and can significantly improve runtime performance.
\section{Acknowledgements}
This research was supported in part by UConn REP award (KFS: 4648460), the National Science Foundation (NSF) Grants 1743418, NSF 1843025, NSF 1849246 and NSF 1952096. This research is based upon work supported by the U.S. Department of Energy, Office of Energy Efficiency and Renewable Energy (EERE) under the Advanced Manufacturing Office Award Number DE-EE0007613.

\bibliography{anthology,custom,ding,cw}

\begin{thebibliography}{29}
\expandafter\ifx\csname natexlab\endcsname\relax\def\natexlab#1{#1}\fi

\bibitem[{Bonawitz et~al.(2019)Bonawitz, Eichner, Grieskamp, Huba, Ingerman,
  Ivanov, Kiddon, Konecny, Mazzocchi, McMahan et~al.}]{bonawitz2019towards}
Keith Bonawitz, Hubert Eichner, Wolfgang Grieskamp, Dzmitry Huba, Alex
  Ingerman, Vladimir Ivanov, Chloe Kiddon, Jakub Konecny, Stefano Mazzocchi,
  H~Brendan McMahan, et~al. 2019.
\newblock Towards federated learning at scale: System design.
\newblock \emph{arXiv preprint arXiv:1902.01046}.

\bibitem[{Bonawitz et~al.(2017)Bonawitz, Ivanov, Kreuter, Marcedone, McMahan,
  Patel, Ramage, Segal, and Seth}]{bonawitz2017practical}
Keith Bonawitz, Vladimir Ivanov, Ben Kreuter, Antonio Marcedone, H~Brendan
  McMahan, Sarvar Patel, Daniel Ramage, Aaron Segal, and Karn Seth. 2017.
\newblock Practical secure aggregation for privacy-preserving machine learning.
\newblock In \emph{ACM CCS}, pages 1175--1191.

\bibitem[{Deng et~al.(2021)Deng, Wang, Li, Shang, Liu, Rajasekaran, and
  Ding}]{deng2021tag}
Jieren Deng, Yijue Wang, Ji~Li, Chao Shang, Hang Liu, Sanguthevar Rajasekaran,
  and Caiwen Ding, editors. 2021.
\newblock \emph{Findings of the Association for Computational Linguistics:
  EMNLP 2021}. Association for Computational Linguistics.

\bibitem[{Ding et~al.(2017)Ding, Liao, Wang, Li, Liu, Zhuo, Wang, Qian, Bai,
  Yuan, Ma, Zhang, Tang, Qiu, Lin, and Yuan}]{10.1145/3123939.3124552}
Caiwen Ding, Siyu Liao, Yanzhi Wang, Zhe Li, Ning Liu, Youwei Zhuo, Chao Wang,
  Xuehai Qian, Yu~Bai, Geng Yuan, Xiaolong Ma, Yipeng Zhang, Jian Tang, Qinru
  Qiu, Xue Lin, and Bo~Yuan. 2017.
\newblock \href {https://doi.org/10.1145/3123939.3124552} {Circnn: Accelerating
  and compressing deep neural networks using block-circulant weight matrices}.
\newblock In \emph{Proceedings of the 50th Annual IEEE/ACM International
  Symposium on Microarchitecture}, MICRO-50 '17, page 395–408, New York, NY,
  USA. Association for Computing Machinery.

\bibitem[{Dwork(2008)}]{10.5555/1791834.1791836}
Cynthia Dwork. 2008.
\newblock Differential privacy: A survey of results.
\newblock In \emph{Proceedings of the 5th International Conference on Theory
  and Applications of Models of Computation}, TAMC'08, page 1–19, Berlin,
  Heidelberg. Springer-Verlag.

\bibitem[{Dwork et~al.(2014)Dwork, Roth et~al.}]{dwork2014algorithmic}
Cynthia Dwork, Aaron Roth, et~al. 2014.
\newblock The algorithmic foundations of differential privacy.
\newblock \emph{Foundations and Trends in Theoretical Computer Science},
  9(3-4):211--407.

\bibitem[{Geyer et~al.(2017)Geyer, Klein, and Nabi}]{geyer2017differentially}
Robin~C Geyer, Tassilo Klein, and Moin Nabi. 2017.
\newblock Differentially private federated learning: A client level
  perspective.
\newblock \emph{arXiv preprint arXiv:1712.07557}.

\bibitem[{Gui et~al.(2019)Gui, Wang, Yang, Yu, Wang, and Liu}]{gui2019model}
Shupeng Gui, Haotao~N Wang, Haichuan Yang, Chen Yu, Zhangyang Wang, and Ji~Liu.
  2019.
\newblock Model compression with adversarial robustness: A unified optimization
  framework.
\newblock In \emph{NeurIPS}, pages 1285--1296.

\bibitem[{Gurevin et~al.(2021)Gurevin, Bragin, Ding, Zhou, Pepin, Li, and
  Miao}]{Gurevin2021Enabling}
Deniz Gurevin, Mikhail Bragin, Caiwen Ding, Shanglin Zhou, Lynn Pepin, Bingbing
  Li, and Fei Miao. 2021.
\newblock \href {https://doi.org/10.24963/ijcai.2021/344} {Enabling
  retrain-free deep neural network pruning using surrogate lagrangian
  relaxation}.
\newblock In \emph{Proceedings of the Thirtieth International Joint Conference
  on Artificial Intelligence, {IJCAI-21}}, pages 2497--2504. International
  Joint Conferences on Artificial Intelligence Organization.
\newblock Main Track.

\bibitem[{Hochreiter and Schmidhuber(1997)}]{HochSchm97}
Sepp Hochreiter and Jürgen Schmidhuber. 1997.
\newblock Long short-term memory.
\newblock \emph{Neural Computation}, 9(8):1735--1780.

\bibitem[{Lin et~al.(2020)Lin, Wang, Li, Deng, Wang, and Ding}]{lin2020esmfl}
Sheng Lin, Chenghong Wang, Hongjia Li, Jieren Deng, Yanzhi Wang, and Caiwen
  Ding. 2020.
\newblock Esmfl: Efficient and secure models for federated learning.
\newblock \emph{arXiv preprint arXiv:2009.01867}.

\bibitem[{Lueker(1998)}]{lueker1998exponentially}
George~S Lueker. 1998.
\newblock Exponentially small bounds on the expected optimum of the partition
  and subset sum problems.
\newblock \emph{Random Structures \& Algorithms}, 12(1):51--62.

\bibitem[{Ma et~al.(2020)Ma, Guo, Niu, Lin, Tang, Ma, Ren, and
  Wang}]{ma2020pconv}
Xiaolong Ma, Fu-Ming Guo, Wei Niu, Xue Lin, Jian Tang, Kaisheng Ma, Bin Ren,
  and Yanzhi Wang. 2020.
\newblock Pconv: The missing but desirable sparsity in dnn weight pruning for
  real-time execution on mobile devices.
\newblock In \emph{AAAI}, pages 5117--5124.

\bibitem[{McMahan et~al.(2017{\natexlab{a}})McMahan, Moore, Ramage, Hampson,
  and y~Arcas}]{google-fl}
H.~Brendan McMahan, Eider Moore, Daniel Ramage, Seth Hampson, and Blaise~Aguera
  y~Arcas. 2017{\natexlab{a}}.
\newblock \href {http://arxiv.org/abs/1602.05629} {Communication-efficient
  learning of deep networks from decentralized data}.
\newblock In \emph{(AISTATS)}.

\bibitem[{McMahan et~al.(2017{\natexlab{b}})McMahan, Moore, Ramage, Hampson,
  and y~Arcas}]{mcmahan2017communicationefficient}
H.~Brendan McMahan, Eider Moore, Daniel Ramage, Seth Hampson, and
  Blaise~Agüera y~Arcas. 2017{\natexlab{b}}.
\newblock \href {http://arxiv.org/abs/1602.05629} {Communication-efficient
  learning of deep networks from decentralized data}.

\bibitem[{Merity et~al.(2016)Merity, Xiong, Bradbury, and
  Socher}]{merity2016pointer}
Stephen Merity, Caiming Xiong, James Bradbury, and Richard Socher. 2016.
\newblock Pointer sentinel mixture models.
\newblock \emph{arXiv preprint arXiv:1609.07843}.

\bibitem[{Papernot et~al.(2018)Papernot, Song, Mironov, Raghunathan, Talwar,
  and Erlingsson}]{conf/iclr/PapernotSMRTE18}
Nicolas Papernot, Shuang Song, Ilya Mironov, Ananth Raghunathan, Kunal Talwar,
  and {\'{U}}lfar Erlingsson. 2018.
\newblock Scalable private learning with {PATE}.
\newblock In \emph{{ICLR} 2018,}.

\bibitem[{Peter et~al.(2012)Peter, Kronberg, Trei, and
  Katzenbeisser}]{Peter2012AdditivelyHE}
Andreas Peter, Max Kronberg, Wilke Trei, and Stefan Katzenbeisser. 2012.
\newblock Additively homomorphic encryption with a double decryption mechanism,
  revisited.
\newblock In \emph{ISC}.

\bibitem[{Ren et~al.(2020)Ren, Zhang, Wang, Lin, Dong, Chen, Xie, and
  Wang}]{ren2020darb}
Ao~Ren, Tao Zhang, Yuhao Wang, Sheng Lin, Peiyan Dong, Yen-Kuang Chen, Yuan
  Xie, and Yanzhi Wang. 2020.
\newblock Darb: A density-adaptive regular-block pruning for deep neural
  networks.
\newblock In \emph{AAAI}, pages 5495--5502.

\bibitem[{Truex et~al.(2019)Truex, Baracaldo, Anwar, Steinke, Ludwig, Zhang,
  and Zhou}]{truex2019hybrid}
Stacey Truex, Nathalie Baracaldo, Ali Anwar, Thomas Steinke, Heiko Ludwig, Rui
  Zhang, and Yi~Zhou. 2019.
\newblock A hybrid approach to privacy-preserving federated learning.
\newblock In \emph{Proceedings of the 12th ACM Workshop on Artificial
  Intelligence and Security}, pages 1--11.

\bibitem[{Vaswani et~al.(2017)Vaswani, Shazeer, Parmar, Uszkoreit, Jones,
  Gomez, Kaiser, and Polosukhin}]{vaswani2017attention}
Ashish Vaswani, Noam Shazeer, Niki Parmar, Jakob Uszkoreit, Llion Jones,
  Aidan~N Gomez, {\L}ukasz Kaiser, and Illia Polosukhin. 2017.
\newblock Attention is all you need.
\newblock In \emph{Advances in neural information processing systems}, pages
  5998--6008.

\bibitem[{Wang et~al.(2020)Wang, Deng, Guo, Wang, Meng, Liu, Ding, and
  Rajasekaran}]{wang2020sapag}
Yijue Wang, Jieren Deng, Dan Guo, Chenghong Wang, Xianrui Meng, Hang Liu,
  Caiwen Ding, and Sanguthevar Rajasekaran. 2020.
\newblock Sapag: a self-adaptive privacy attack from gradients.
\newblock \emph{arXiv preprint arXiv:2009.06228}.

\bibitem[{Wang et~al.(2021)Wang, Wang, Wang, Zhou, Liu, Bi, Ding, and
  Rajasekaran}]{ijcai2021-432}
Yijue Wang, Chenghong Wang, Zigeng Wang, Shanglin Zhou, Hang Liu, Jinbo Bi,
  Caiwen Ding, and Sanguthevar Rajasekaran. 2021.
\newblock \href {https://doi.org/10.24963/ijcai.2021/432} {Against membership
  inference attack: Pruning is all you need}.
\newblock In \emph{Proceedings of the Thirtieth International Joint Conference
  on Artificial Intelligence, {IJCAI-21}}, pages 3141--3147. International
  Joint Conferences on Artificial Intelligence Organization.
\newblock Main Track.

\bibitem[{Wen et~al.(2016)Wen, Wu, Wang, Chen, and Li}]{wen2016learning}
Wei Wen, Chunpeng Wu, Yandan Wang, Yiran Chen, and Hai Li. 2016.
\newblock Learning structured sparsity in deep neural networks.
\newblock In \emph{Advances in Neural Information Processing Systems}, pages
  2074--2082.

\bibitem[{Wu et~al.(2021)Wu, Zheng, Dou, Chen, Deng, Chen, Xu, Gao, Li, Wang
  et~al.}]{wu2021novel}
Xin Wu, Hao Zheng, Zuochao Dou, Feng Chen, Jieren Deng, Xiang Chen, Shengqian
  Xu, Guanmin Gao, Mengmeng Li, Zhen Wang, et~al. 2021.
\newblock A novel privacy-preserving federated genome-wide association study
  framework and its application in identifying potential risk variants in
  ankylosing spondylitis.
\newblock \emph{Briefings in Bioinformatics}, 22(3):bbaa090.

\bibitem[{Yao(1986)}]{Yao}
A.~C. Yao. 1986.
\newblock \href {https://doi.org/10.1109/SFCS.1986.25} {How to generate and
  exchange secrets}.
\newblock In \emph{27th Annual Symposium on Foundations of Computer Science
  (sfcs 1986)}, pages 162--167.

\bibitem[{Yuan et~al.(2021)Yuan, Behnam, Cai, Shafiee, Fu, Liao, Li, Ma, Deng,
  Wang, Bojnordi, Wang, and Ding}]{9474235}
Geng Yuan, Payman Behnam, Yuxuan Cai, Ali Shafiee, Jingyan Fu, Zhiheng Liao,
  Zhengang Li, Xiaolong Ma, Jieren Deng, Jinhui Wang, Mahdi Bojnordi, Yanzhi
  Wang, and Caiwen Ding. 2021.
\newblock \href {https://doi.org/10.23919/DATE51398.2021.9474235} {Tinyadc:
  Peripheral circuit-aware weight pruning framework for mixed-signal dnn
  accelerators}.
\newblock In \emph{2021 Design, Automation Test in Europe Conference Exhibition
  (DATE)}, pages 926--931.

\bibitem[{Zhao et~al.(2017)Zhao, Liao, Wang, Li, Tang, and
  Yuan}]{zhao2017theoretical}
Liang Zhao, Siyu Liao, Yanzhi Wang, Zhe Li, Jian Tang, and Bo~Yuan. 2017.
\newblock Theoretical properties for neural networks with weight matrices of
  low displacement rank.
\newblock In \emph{International Conference on Machine Learning}, pages
  4082--4090.

\bibitem[{Zhu et~al.(2019)Zhu, Liu, and Han}]{zhu2019deep}
Ligeng Zhu, Zhijian Liu, and Song Han. 2019.
\newblock Deep leakage from gradients.
\newblock In \emph{NeurIPS}, pages 14774--14784.

\end{thebibliography}
\bibliographystyle{acl_natbib}




\end{document}


\maketitle

\appendix
\section{Security Model}\label{sec:semodel}
In our system, we assume a semi-honest (or \emph{honest-but-curious}) adversary $\mathcal{A}$
who is able to compromise a subset of the clients. Moreover, the maximum number of clients that $\mathcal{A}$ is able to compromise is bounded by a threshold $T$. There are no restrictions on collusion among clients and the servers, {i.e.,} $\mathcal{A}$ can control a set of clients and one server at the same time. Such an adversary is called an ``admissible adversary'' as mentioned by Mohassel and Zhang~\cite{mohassel2017secureml}. 
For the non-colluding servers \CSP and \AS, we also consider the semi-honest adversary model.

At a high level, we want to provide a security guarantee to the uncorrupted parties against 
an admissible adversary $\mathcal{A}$ who might compromise a subset of clients. 
Therefore, the desired security is that no {\it probabilistic polynomial time} (ppt) adversary 
can learn anything about the data beyond corrupted clients' data. 
Known algorithms that achieve there security properties are not efficient enough for practical use. Usually, one allows for some form of leakage to the server. 
As initiated in the work of Curtmola \etal\cite{CurtmolaGKO06}
and Chase and Kamara~\cite{chase2010structured}, one can define acceptable \emph{leakage functions} and devise an efficient protocol that \emph{provably} leak only the values of
these functions.

In what follows, we define the security model of our proposed approach in details. We start with the definition of ideal functionality our FL framework in Fig.~\ref{fig:func}. Then we formally describe our security models.\\

\begin{figure}[h]
    \centering
    \noindent\fbox{%
    \parbox{0.45\textwidth}{%
       {\bf Parameters:} Clients $\cl_1, \cl_2, ..., \cl_m$, two servers $\mathcal{S}_0, \mathcal{S}_1$ and security parameter $\lambda$.\\
       {\bf Uploading Data:} Each client $\cl_i$ inputs the disseminated data (internally stored) $d_i$.\\
       {\bf Federated Training:} All clients and $\mathcal{S}_0$, $\mathcal{S}_1$ agree on an ML algorithm $f$.
       $\mathcal{S}_0$ and $\mathcal{S}_1$ compute $ y \larr f(d_1, d_2, ..., d_m)$, and $\mathcal{S}_0$ sends $y$ to each $\cl_i$, output a leakage function $\mathcal{L}$. 
     }%
    }
    \caption{Ideal Machine Learning Functionality $\mathcal{F}_{\class{ML}}$}\label{fig:func}
\end{figure}

\begin{definition}\label{def:security}
Let $\pi$ be an secure FL protocol and  a set of clients $\mathcal{C} = \{\cl_1, \cl_2, ..., \cl_m\}$ participate the execution of $\pi$.
Consider the following probabilistic experiments where $\mathcal{Z}$ is an environment, 
$\mathcal{A}$ is ppt semi-honest adversary, $\mathcal{S}$ is a simulator, 
and $\mathcal{L}$ is a stateful leakage function. 

\begin{itemize}
    \item {\bf Real}$_{\mathcal{A}}^{\pi}(1^{\lambda})$: 
    $\mathcal{Z}$ outputs a dataset $\mathcal{D}$, horizontally partition $\mathcal{D}$, and distributes partitions to clients. All parties then execute protocol $\pi$ on the given security parameter $\lambda$. 
    During the execution of $\pi$, corrupted parties forward their protocol inputs, outputs, and internal states to $\mathcal{A}$. $\mathcal{A}$ finally outputs a bit $b$ at the end of the experiment.
    
    \item {\bf Ideal}$_{\mathcal{A, S}}^{\pi}(1^{\lambda})$: 
    $\mathcal{Z}$ outputs a dataset $\mathcal{D}$ horizontally partition  $\mathcal{D}$, and distributes 
    partitions to clients. Each client submits the received data $d_i$ to the ideal ML functionality $\mathcal{F}_{\class{ML}}$. The $\mathcal{F}_{\class{ML}
    }$ outputs $\mathcal{L}$ and forwards it to $\mathcal{S}$. 
    Given $\mathcal{L}$ and $1^{\lambda}$, $\mathcal{S}$ generates the outputs and reveal it to $\mathcal{A}$. $\mathcal{A}$ finally outputs a bit $b$ at the end of the experiment.
\end{itemize}

We say that protocol $\pi$ is $\mathcal{L}$-semantically secure if 
for all $\mathcal{Z}$, all ppt $\mathcal{A}$, there exists a ppt simulator $\mathcal{S}$ such that
\[
\Big|\Pr\big[{\bf Real}_{\mathcal{A}}^{\pi}(1^{\lambda}) = 1\big] - \Pr\big[{\bf Ideal}_{\mathcal{A, \textsf{S}}}^{\pi}(1^{\lambda})\big] = 1 \Big| = \class{negl}(\lambda).
\]
\end{definition}

The definition above captures the security guarantees that $\mathcal{A}$ learns nothing beyond allowed
leakage function $\mathcal{L}$ and the corrupted parties' data. 
Intuitively, in the ideal interaction, parties simply forward the inputs they receive to an incorruptible functionality and forward the functionality's response to the environment. The ideal functionality perform the computation on behalf the parties. 
Given $\mathcal{L}$, the simulator needs to carry out the same outcome in the ideal interaction that the adversary achieves in the real interactions.

Next, we introduce the definition of differentially private secure federated learning protocol. 
Consider that $\mathcal{A}$ has dataset $\mathcal{D}$ that is generated by $\mathcal{Z}$ and 
repeatedly run the ideal experiment with different $\mathcal{D}$s. 

 \begin{definition}[differentially private secure FL] \label{def:sdpfl}
Let $\pi$ be a $\mathcal{L}$-semantically secure FL protocol with a leakage function $\mathcal{L}$. 
$\pi$ is said to be a $\epsilon$-deferentially private secure FL ($\epsilon$-DP secure FL) 
iff, for any two neighboring $\mathcal{D,D'}$, for any adversary (even computational unbounded) $\mathcal{A}$, we have:
\[
\Pr\big[\textup{View}_{\mathcal{A}}({\bf Ideal}_{\mathcal{A, S}}^{\pi}, \mathcal{D})\big] \leq e^{\epsilon} \Pr\big[\textup{View}_{\mathcal{A}}({\bf Ideal}_{\mathcal{A, S}}^{\pi}, \mathcal{D'})\big] + \delta
\]
where $\textup{View}_{\mathcal{A}}({\bf Ideal}_{\mathcal{A, S}}^{\pi}, \mathcal{D})$ denotes the view of $\mathcal{A}$ running the ideal experiment using data $\mathcal{D}$.
\end{definition}

Definition~\ref{def:sdpfl} ensures that by running the defined FL protocol, the information related to any single record in the entire training dataset is bounded in a differentially private sense. 


\eat{
\subsection{Security and Privacy Semantics}
In this section, we discuss the semantics of privacy and security ensured by Definition \ref{def:sdpfl}. The security and privacy provided by $\epsilon$-SDPFL can be interpreted in terms of: 1) computationally secure with allowed leakage and 2) plausible deniability on information of joint data. The first claims holds followed by $\mathcal{L}-$semantically secure definition, which indicates that as long as the number of honest party does not fall below the threshold, the adversary learns nothing about the honest parties' internal stored data beyond the information revealed in the allowed leakage. The plausible deniability on information of joint data is ensured by reveal leakage that is protected by differential privacy mechanism. Thus the leakage is then information theoretically secure, namely any (even computational unbounded) adversary will not able to distinguish if certain record is in or absence from the joint data hosted by the honest parties.
}

\section{\system Security Sketch}\label{sec:secsk}
\eat{
We consider security and privacy guarantee for set of data owners against a pair of semi-honest but non-colluding servers \AS and \CSP(\AS and \CSP follow the exact predefined protocol and never deviate from the protocol arbitrarily). According to our framework design, at epoch $t$, the view of \AS consists of a set of encrypted updates $\{ \cup_{i \in L_t} ({\bf \tilde{\Delta {\bf{W}}}^{(t)}_i }) \}$, a set of encrypted Gaussian noise $\{\cup_{i \in L_t} ({\bf \tilde{n}_i})\}$, and both the plaintext and ciphertext of the updated global model $\tilde {\bf{W}}^{(t)},  {\bf{W}}^{(t)}$, respectively. The $\CSP$'s view for each epoch only consists of $\tilde {\bf{W}}^{(t)}$ and $  {\bf{W}}^{(t)}$. \eat{For Analyst, we assume it only be able to obtain the final released model $ {\bf{W}}^{(T)}$.}\textcolor{red}{(Explain Assumption?)} As $\{ \cup_{i \in L_t} ({ \tilde{\Delta {\bf{W}}}^{(t)}_i }) \}$, $\{\cup_{i \in L_t} ({\tilde{n}_i})\}$ and $\tilde {\bf{W}}^{(t)}$ reveals nothing other than negligible information bounded by security parameter $\kappa$ due to the semantic security provided by paillier cryptosystem\cw{reference}. Thus, besides the intermediate global model, all the other information obtained by \AS and \CSP are considered semantically secure. Moreover, as the aggregated model at each epoch has perturbed by Gaussian noises, thus the intermediate global model $ {\bf{W}}^{(t)}$ satisfy $(O(q\epsilon), \delta)$-DP \cw{reference}, where $q$ is the fraction of group size for current epoch over the total number of data owners. Thus combine with the composition theorem, we can have the statement that after training completed, the information that $\AS$ and $\CSP$ obtained are either semantically secure or deferentially private.
}
Followed by the security model, in this section, we provide sketch security proof of our proposed \system
framework. We start with the following lemma:
\begin{lemma}\label{theorem:sensitivity} Given a $L$ local updates $\{{\bf \Delta {\bf{W}}}^{(t)}_{1},  {\bf \Delta{\bf{W}}}^{(t)}_{2}, ..., {\bf \Delta {\bf{W}}}^{(t)}_{L}\}$, which are clipped by a bound of $C$. Given $\mathcal{T}$ denote the BHM compression with compression ratio $\gamma$ and scaling factor $\kappa$. Then the sensitivity of ${\bf \Delta {\bf{w}}}^{(t)} \gets \frac{1}{L}\sum_{i=1}^{L}\mathcal{T}({\bf \Delta {\bf{W}}}^{(t)}_i, \gamma, \kappa)$ is bounded by $\kappa C$.
\end{lemma}
\vspace{-3mm}
\begin{proof}
For each local model update, as ${\bf \Delta {\bf{W}}}^{(t)}_i$ is clipped by a bound of $C$, thus it's sensitivity is then bounded by $C$~\cite{abadi2016deep}.  The compression mechanism scales each element in {${\bf \Delta {\bf{W}}}^{(t)}_i$} with the scaling factor $\kappa$, thus the sensitivity of {$\mathcal{T}({\bf \Delta {\bf{W}}}^{(t)}_i, \gamma, \kappa)$}, denoted as $S({\bf \Delta {\bf{w}}}_i)$ is thus bounded by $\kappa C$. In addition, $S({\bf \Delta {\bf{w}}}) \leq \max_i S({\bf \Delta {\bf{w}}}_i) = {\kappa}C$, thus the lemma holds.
\end{proof}
\begin{lemma}[Privacy Property of Gaussian Mechanism]\label{theorem:dp}
Let $\mathcal{M}$ (a probabilistic mechanism) as a Gaussian mechanism denoted as {$\mathcal{M} = f(\mathcal{D}) + \mathcal{N}\left( 0, ~\sigma^2S_{f}^2 \right)$, where $S_{f}^2$} is the sensitivity of $f$. Given privacy parameter $\epsilon$ and $\delta$, and let {$\sigma = \sqrt{2\log\frac{1.25}{\delta}}/\epsilon$}. Then $\mathcal{M}$ satisfies  $(\epsilon, \delta)$-DP.
\end{lemma}

\begin{theorem}
The proposed \system satisfies Definition~\ref{sec:se}
\end{theorem}
\begin{proof}
Let $\mathcal{M}$ be a single training round of \sytstem, and let $\class{DS}$ be the union set of all clients' data. Let {$\mathcal{M}_1 (\class{DS}) = \frac{1}{L}\sum_{i=1}^{L}\mathcal{T}({\bf \Delta {\bf{W}}}^{(t)}_i, \gamma, \kappa)$}, then $\mathcal{M}$ can be expressed as {$\mathcal{M} = \mathcal{T}^{-1}\left(\mathcal{M}_1(\class{DS}) + \mathcal{N}\left( 0, ~\sigma^2\kappa^2C^2 \right)\right)$.} Knowing that $\mathcal{M}_1(\class{DS})$'s sensitivity is $\kappa C$, thus $\mathcal{M}$ is actually a Gaussian mechanism followed by post-processing. Combine Lemma \ref{theorem:sensitivity}, \ref{theorem:dp}, and privacy amplification theorem~\cite{abadi2016deep}, we can conclude that $\mathcal{M}$ satisfies $(\epsilon, \delta)$-DP. In addition, as all local updates are encrypted by Paillier, thus followed by semantic security property~\cite{goldreich2009foundations}, the only leakage of executing a single round of \system should be equal to the output of $\mathcal{M}$. Thus the theorem holds.
\end{proof}

\eat{
\begin{proof}
Given $m$ clients and let the group size is $L$, then the sample ratio is $\frac{L}{m}$. Given any dataset $\mathcal{D}$, the general training mechanism can be expressed as $\mathcal{M} = f(\mathcal{D}) + \mathcal{N}\left( 0, ~\sigma^2{\kappa^2}C^2 \right)$, where $f(\mathcal{D})$ refers to the process of aggregating a global model update. 
According to lemma~\ref{theorem:sensitivity}, we know $f$ has a sensitivity of $\kappa/L\times C$. Combine with the the privacy amplification theorem~\cite{kasiviswanathan2011can, beimel2010bounds}, $\mathcal{M}$ satisfies $(\frac{L}{m}\epsilon, \frac{L}{m}\delta)$-differential privacy.
\end{proof}
}
\section{Convergence Analysis}
The convergence of learning algorithms under differential privacy has been extensively studied. Therefore, in this section, we focus on the theoretical analysis of the convergence properties of the weight pruned ML models.

\subsubsection{Preliminaries and notations} 
We start with introducing related notations that will be used in the following analysis. A list of key notations are presented in in Table \ref{table: notations}. Now, we define the target network $f(x)$ as:
\begin{equation}
    f(x)=W_{n}\sigma (W_{n-1}...(\sigma(W_1 x))
    \label{def_f}
\end{equation}
and we define the original network $g(x)$ as:
\begin{equation}
    g(x)= W_{2n}^g\sigma(W_{2n-1}^g...\sigma(W_1^g(x))
    \label{def_g}
\end{equation}
where $W_i^f$,$W_j^g$ is the randomized weight matrix at $i$-th layer of $f$ and $j$-th layer of $g(x)$. And $\sigma(\cdot)$ is the activation function.\\
A pruned network $\hat{g}(x)$ can be presented as :
\begin{equation}
    \hat{g}(x) = (P_{2n} \odot W_{2n}^g)\sigma(P_{2n-1} \odot W_{2n-1}^g)...\sigma(P_{1} \odot W_{1}^gx)
\end{equation}
Where $P_l$ is the prunning matrix in $l$-th layer.

\begin{table}
\centering
\begin{tabular}{|l|l|} 
\hline
Notation & Definition  \\ 
\hline
$f$              & target function \\
\hline
$g$              & original randomized function \\
\hline
$\hat{g}$              & pruned function \\
\hline
\sigma        &  activation function           \\ 
\hline
$w_i,\mathbf{w}_i$,$\mathbf{W}_i$        &     the weight value/vector/matrix in $i$th layer        \\ 
\hline
\delta         &    the error probability         \\ 
\hline
\mathbf{p}_i, \mathbf{P}_i        &     the pruning vector/matrix in $i$th layer        \\ 
\hline
 C        &   constant            \\ 
\hline
\epsilon         &     a parameter, \forall \epsilon >0       \\
\hline

\end{tabular}
\caption{Parameter Notations}
\label{table: notations}
\end{table}

\subsubsection{Convergence analysis of Weight Pruned ML models}

Our main goal is to find a pruned network $\hat{g}(x)$ which is competitive with $f(x)$,i,e:
\begin{equation}
 \underset{x \in \chi}{\text{sup}} {\left \| f(x) - \hat{g}(x) \right \|} \leq \epsilon   
\end{equation}

\\
By using our proposed pruning method, the result for finding$\hat{g}(x)$ is fallowing. We prove that from sufficiently logarithmically over-parameterized neural network with random initialized weights there is a subnetwork can match the same performance with high probability.\\
{\bf Theorem 1.} {\it For every network $f$ defined in Eq. \ref{def_f} with depth l and \forall $i \in \{1,2,\dots,n\}$. Consider g defined in Eq.\ref{def_g} is a randomly initialized neural network with 2n layers, and width $poly(d,n,m,1/\epsilon,log1/\delta)$, where $d$ is input size, n is number of layers in $f$, m is the maximum number of neurons in a layer.  The weight initialization distribution belongs to Uniform distribution in range [-1,1]. Then with probability at least $1-\delta$  there is a weight-pruned subnetwork \hat{g} of g such that:} 
\begin{equation}
 \underset{x \in \chi, \left \| W \right \| \leq 1}{\text{sup}} {\left \| f(x) - \hat{g}(x) \right \|} \leq \epsilon   
\end{equation}

In what follows we start with our analysis over simple linear function models with unique variable, then extend the analysis to cover ReLu function, and deep neural networks.\\

\eat{
In the following from section 4.2 to section4.5 , we analysis the objective from simple to complex. We will start from a pruning simple linear function with one variable. Then we consider about pruning a simple ReLU network. Furthermore, we analysis the pruning from a neuron and a layer. Finally,  we give the analysis of our main objective at section 4.6. }

\textbf{Case1. Analysis over Simple linear Network}\\
In this case, $f(x) = w \cdot x$ , and $g(x)=  \left (  \sum_{i=1}^{d}w_i\right )x$.

{\bf Theorem 2.~\cite{lueker1998exponentially}} {\it Let $W^*_1,...,W^*_n$ as $n$ i.i.d. random variables drawn from the Uniform distribution over [-1,1], where $n \geq Clog\frac{2}{\delta}$ ,where $\delta \leq min\{1,\epsilon\}$. Then, with probability at least 1-$\delta$, we have \\
\begin{equation}
\begin{aligned}
    &\exists S \subset \{1,2,...,n\}, \forall W \in [-0.5,0.5],\\
    &s.t \left | W-\sum_{i \in S}W^*_i \right | \leq \epsilon 
\end{aligned}
\end{equation}
}

\textbf{Case 2. Analysis on Simple ReLU Networks}\\
In this case, $f(x) = w \cdot x$, $g(x) = \mathbf{u}\sigma (\mathbf{w}^g x)$ .
because $\sigma$ is ReLU activation function, we have $w =\sigma (w) - \sigma(-w)$. So that the a single ReLU neuron can be written as:

\begin{equation}
\begin{aligned}
x^*  \mapsto \sigma \left ( wx \right ) = \sigma\left ( \sigma(wx)  - \sigma(-wx) \right)
\label{neuron_o} 
\end{aligned}
\end{equation}

On the other hand, this neuron can be present by a two layer network with a pruning matrix $p^*$ for the first layer as:
\begin{equation}
    x^*  \mapsto  \mathbf{u} \sigma \left ( \mathbf{p}\odot  \mathbf{w}^g x\right ) 
    \label{neuron_f}
\end{equation}

we define $\mathbf{w^+}=max\{\mathbf{0},\mathbf{w}\}$,$\mathbf{w^-}=min\{\mathbf{0},\mathbf{w}\}$, $\mathbf{w^+}+\mathbf{w^-}=\mathbf{w}^g$. 
Combine Eq. \ref{neuron_o} and \ref{neuron_f} we have:
\begin{equation}
    x^*  \mapsto  \mathbf{u} \sigma\left( \sigma \left ( \mathbf{p}\odot  \mathbf{w^+}x\right )-\sigma \left ( \mathbf{p}\odot  \mathbf{-w^-}x\right )  \right ) 
\end{equation}

Base on Theorem 2, when $n \geq Clog4/\epsilon$, there exist a pattern of $\mathbf{w}$, such that, with probability $1-\epsilon/2$, 
\begin{equation}
    \forall w^f \in [0,1],  \exists  p \in {0,1}^n, s.t.  \left | w^f- \mathbf{u} \sigma( \mathbf{p} \odot \mathbf{w^+}) \right | < \epsilon/2
    \label{wplus}
\end{equation}

Similarly, we have 

$\mathbf{w}$, such that, with probability $1-\epsilon/2$, 
\begin{equation}
    \forall w^f \in [0,1],  \exists  p \in {0,1}^n, s.t.  \left | w^f- \mathbf{u} \sigma( \mathbf{p} \odot \mathbf{w^-}) \right | < \epsilon/2
    \label{wminus}
\end{equation}
so Consider Eq.\ref{wplus} and \ref{wminus}, we have:
\begin{equation}
\begin{split}
& \text{sup} \left | w^fx- \mathbf{u} \sigma( \mathbf{p} \odot \mathbf{w}x) \right | \\
& \leq \left | \sigma(w^f)x-\sigma(-w^f)x- \mathbf{u} \sigma( \mathbf{p} \odot \mathbf{w^+}x)- \mathbf{u} \sigma( \mathbf{p} \odot \mathbf{w^-}x) \right | \\
& \leq  \text{sup} \left | \sigma (w^f)x- \mathbf{u} \sigma( \mathbf{p} \odot \mathbf{w^+}x) \right |+ \\
& \text{sup} \left | \sigma (w^f)x- \mathbf{u} \sigma( \mathbf{p} \odot \mathbf{w^-}x) \right | \\
& \leq \epsilon/2 + \epsilon/2\\ 
& \leq \epsilon
\end{split}
\label{conv_relu}
\end{equation}

\textbf{Case 3. Analysis on a Neuron}\\
\begin{equation}
\begin{split}
&\text{sup}
\left | \mathbf{w}^f\mathbf{x}- \mathbf{u} \sigma( \mathbf{p} \odot \mathbf{wx}) \right | \\
&\leq \text{sup} \left | \sum_{i=1}^{m}\left (  w_i^fx_i- \mathbf{u}_i \sigma( \mathbf{p}_i \odot \mathbf{w}_ix_i)\right ) \right | \\ 
&\leq \text{sup}\sum_{i=1}^{m} \left | w_i^fx_i- \mathbf{u}_i \sigma( \mathbf{p}_i \odot \mathbf{w}_ix_i) \right | \\ 
&\leq \sum_{i=1}^{m} \text{sup} \left | w_i^fx_i- \mathbf{u}_i \sigma( \mathbf{p}_i \odot \mathbf{w}_ix_i) \right | \\ 
&\leq  m \cdot  \frac{\epsilon}{m}\\
&\leq \epsilon
\end{split}
\label{conv_neuron}
\end{equation}

\textbf{Case 4. Analysis on a single network Layer}\\
\begin{equation}
\begin{split}
&\text{sup}
\left | \mathbf{W}^f\mathbf{x}- \mathbf{u} \sigma( \mathbf{p} \odot \mathbf{W}^g\mathbf{x}) \right | \\
&\leq \text{sup} \left | \sum_{j=1}^{k}\sum_{i=1}^{m}\left (  w_{j,i}^fx_i- \mathbf{u}_i \sigma( \mathbf{p}_{j,i} \odot \mathbf{w}_{j,i}x_i)\right ) \right | \\ 
&\leq \text{sup}\sum_{j=1}^{k}\sum_{i=1}^{m} \left | w_{j,i}^fx_i- \mathbf{u}_i \sigma( \mathbf{p}_{j,i} \odot \mathbf{w}_{j,i}x_i) \right | \\ 
&\leq \sum_{j=1}^{k}\sum_{i=1}^{m} \text{sup} \left | w_{j,i}^fx_i- \mathbf{u}_i \sigma( \mathbf{p}_{j,i} \odot \mathbf{w}_{j,i}x_i) \right | \\ 
&\leq  k \cdot m  \cdot \frac{\epsilon}{mk}\\
&\leq \epsilon
\end{split}
\label{conv_layer}
\end{equation}

\textbf{Case 5. Putting it all together}\\
We now provide the general case analysis, where $f(x)$ is defined as Eq.\ref{def_f},  $g(x)$ is defined as Eq.\ref{def_g}. With probability over $1-\epsilon$, we can conclude that:
\begin{equation}
\begin{split}
& \left \| f(x)-\hat g(x) \right \| \\ 
& =\left \| \mathbf{W}_n \mathbf{x}_n - \mathbf{P}_{2n} \odot \mathbf{W}^g_{2n}\mathbf{x}^g_n \sigma(\mathbf{P}_{2n-1} \odot \mathbf{x}^g_{2n-1}) \right \|  \\
& \leq  \left \| \mathbf{W}_n\mathbf{x}_n -  \mathbf{W}_n\mathbf{x}_n^g \right \| + \\ 
& \left \| \mathbf{W}_n\mathbf{x}_n^g -  \mathbf{P}_{2n} \odot \mathbf{W}^g_{2n}\mathbf{x}^g_n \sigma(\mathbf{P}_{2n-1} \odot \mathbf{x}^g_{2n-1}) \right \|  \\
& \leq \left \| \mathbf{x}_n-\mathbf{x}_n ^g \right \| + \\
& \left \| \mathbf{W}_n\mathbf{x}_n^g -  \mathbf{P}_{2n} \odot \mathbf{W}^g_{2n}\mathbf{x}^g_n \sigma(\mathbf{P}_{2n-1} \odot \mathbf{x}^g_{2n-1}) \right \|  \\
& \leq \epsilon/2+ \epsilon/2\\
& \leq \epsilon
\end{split}
\end{equation}

\section{\system's Correctness Analysis}
The correctness of the aggregation follows directly from the properties of 
the AHE, our use of compression, DP, 
and our two-party computation protocol. 
First $\AS$ first gets the aggregation from the homomorphic summation, 
i.e., ${\bf c}^{(t)} = \AHE.\enc((\sum_i^L \hat{\Delta {\bf{W}}}^{(t)}_{i}) / L)$.
As each user multiplies the local update by a factor of $\tfrac{1}{L}$, 
the aggregated model update summarized above is actually the average of all encrypted local updates in their compressed form.
During Secure Aggregation phase, \AS and \CSP execute so that \AS can finally obtain a differentially private aggregation $\tilde{{\Delta {\bf{W}}}}^{(t)}$.
We use a ``masking trick" that simplifies the protocol's complexity. The protocol is designed to have three phases: 
\begin{enumerate}
\item \AS randomly samples a noise vector $\vect{v}$, encrypts it, and homomorphically adds it to the aggregated global model update;
\item \AS sends ${\bf \hat{n}} = \Enc(\pk, {\bf v}) \oplus {\bf c}$ to \CSP. Note that ${\bf c}$ is the encryption of the aggregated model update ${\Delta {\bf{W}}}^{(t)}$;
\item \CSP decrypts ${\bf n} = \AHE.\dec(\sk, {\bf \hat{n}})$.
\CSP and \AS execute a secure 2PC protocol~\cite{goldreich2009foundations}.
\end{enumerate}

Now, after the decryption, \CSP holds input $\vect{n}  = {\Delta {\bf{W}}}^{(t)} + \vect{v}$, it randomly generates a binary vector $\vect{b}_\CSP$. \AS has the masking vector $\vect{v}$, 
it also generates binary vector $\vect{b}_\AS$. 
 \AS generates a GC that first transforms
 $\vect{b}_\AS + \vect{b}_\CSP$ to ${\bf b_G}$,
 a Gaussian noise $\mathcal{N}\left( 0, ~\sigma^2{\kappa^2}C^2 \right)$ using fundamental transformation law, then securely computes $\tilde{{\Delta {\bf{W}}}}^{(t)} = {\Delta {\bf{W}}}^{(t)} + {\bf b_G}$. 
In particular, our 2PC circuits securely computes:
\begin{align*}
\small
\tilde{{\Delta {\bf{W}}}}^{(t)} & = {\Delta {\bf{W}}}^{(t)} + {\bf b_G}\\
& = ({\underbrace{%
       {\Delta {\bf{W}}}^{(t)} + \vect{v} + \vect{b}_\CSP
    }_{\text{\CSP's inputs}}} )
    + (
    {\underbrace{%
     - \vect{v} + \vect{b}_\AS }_{\text{\AS's inputs}}})
\end{align*}
\AS finally gets the aggregated $\tilde{{\Delta {\bf{W}}}}$ with Gaussian noise at the end of the execution.

The security of our protocol follows the security and privacy properties from AHE, 2PC, and our usage Gaussion noises. It is not hard to see that our protocol satisfies Defintion~\ref{def:sdpfl}. Due to space limit, we defer
the formal proof to the full version of this paper.
\bibliography{ref, ding, reference, cw}